\documentclass{PoS}

\title{Photonuclear vector meson production in ultra-peripheral Pb-Pb
collisions studied by the ALICE experiment at the LHC}

\ShortTitle{Photonuclear vector meson production in ultra-peripheral Pb-Pb
collisions}

\author{\speaker{Joakim Nystrand} (for the ALICE Collaboration) \\
        Department of Physics and Technology, 
        University of Bergen, 
        Bergen, Norway \\
        E-mail: \email{Joakim.Nystrand@ift.uib.no}}

\abstract{The strong electromagnetic fields surrounding the Pb-ions accelerated
at the CERN Large Hadron Collider (LHC) allow two-photon and
photonuclear interactions to be studied in a so far unexplored
kinematic regime. Exclusive photoproduction of vector mesons can be
studied in ultra-peripheral collisions, where the impact parameters
are larger than the sum of the nuclear radii and hadronic interactions
are strongly suppressed.

During the heavy-ion runs at the LHC in 2010 and 2011, the ALICE
collaboration used special triggers to select ultra-peripheral collisions. 
These triggers were based on the Muon spectrometer, the Time-of-Flight 
detector, the Silicon Pixel detector, and the VZERO scintillator array. 
Information from other detectors was also used in the analysis. The cross 
section for coherent photoproduction of $J/\psi$ mesons at forward rapidities 
will be presented. The result will be compared to model calculations and 
its implications for nuclear gluon shadowing will be discussed.}

\FullConference{Xth Quark Confinement and the Hadron Spectrum,\\
		October 8-12, 2012\\
		TUM Campus Garching, Munich, Germany}

\begin{document}

\section{Introduction}

Collisions between heavy-ions at the CERN Large Hadron Collider (LHC) can be utilized to 
study particle production in photonuclear and two-photon interactions~\cite{Review2008,Review2005}. 
These interactions may occur in ultra-peripheral 
collisions with impact parameters of several tens or even hundreds of femtometers, where 
the background from hadronic processes is negligible. In this talk, results on coherent 
photoproduction of $J/\psi$ vector mesons and two-photon production of $\mu$--pairs 
measured by the ALICE Collaboration at the LHC will be presented~\cite{Abelev:2012ba}. 

Following the first calculation more than 10 years ago~\cite{starlight}, 
exclusive photoproduction of vector mesons in heavy-ion interactions has attracted an 
increased theoretical interest in recent 
years~\cite{Lappi:2013am,Rebyakova:2011vf,Adeluyi:2012ph,Cisek:2012yt,Goncalves:2011vf}. 
The first results from the Brookhaven Relativistic Heavy-Ion 
Collider showed the experimental feasibility of these studies~\cite{Adler:2002sc,Afanasiev:2009hy}, and  
the increased energy at the LHC leads to significantly higher cross sections 
for heavy vector mesons. The higher energy also means that partons at lower Bjorken-x are probed. 

\begin{figure}
\includegraphics[height=5.0cm]{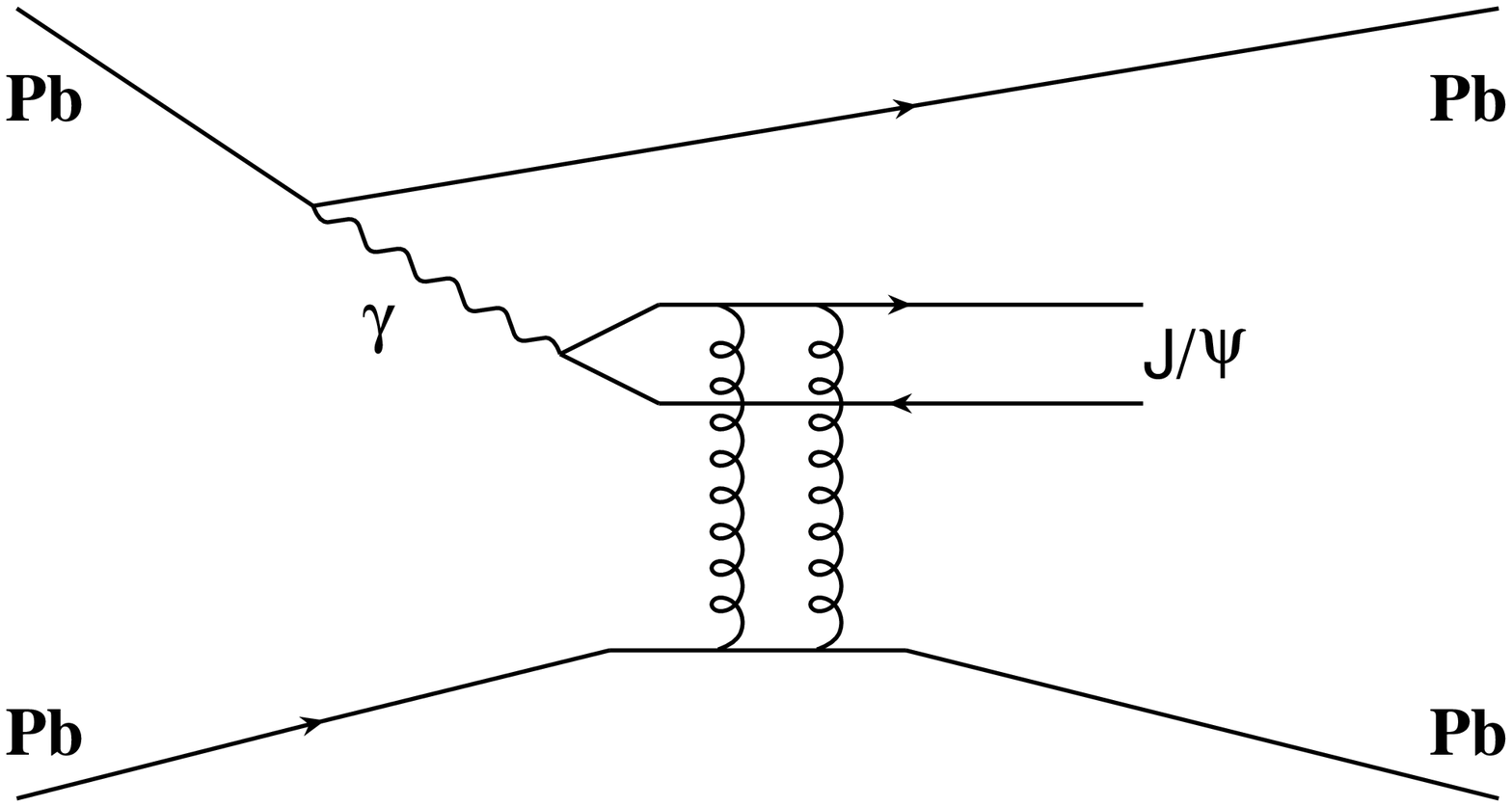}
\includegraphics[height=5.0cm]{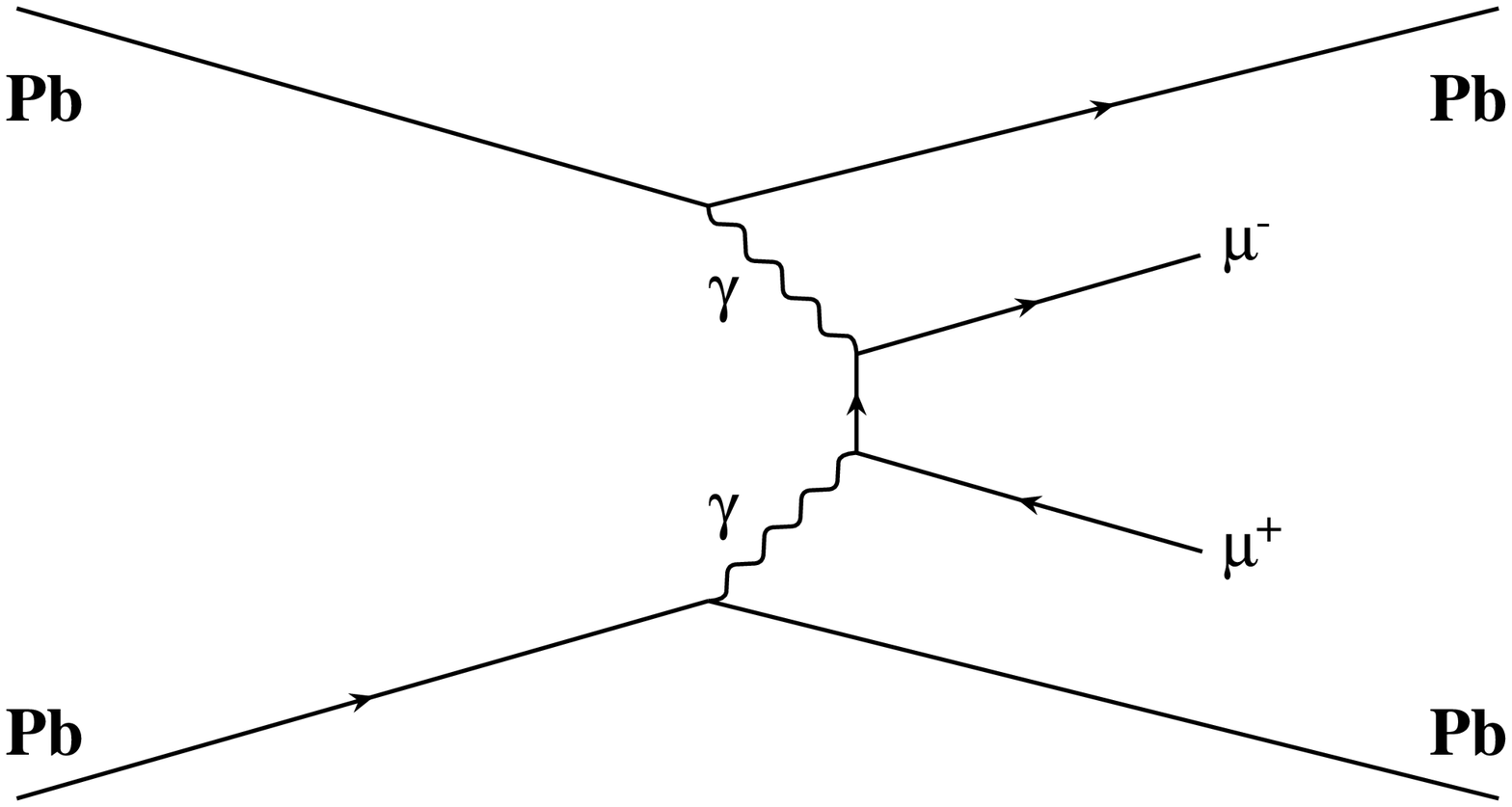}
\caption{Examples of Feynman diagrams for exclusive $J/\psi$ production (left) and two-photon 
production of dimuon pairs (right).}
\label{fig1}
\end{figure}

Exclusive photonuclear production of vector mesons implies that there is no net color 
transfer between the photon and the nucleus. The interaction can proceed via the exchange 
of two gluons, as indicated by the Feynman diagram in Fig.~\ref{fig1} (left). The events considered
here are thus characterized by a single $J/\psi$ meson but no other particles being produced. 
The $J/\psi$ is studied through its dimuon decay. 
Another process with a similar topology is two-photon production of dilepton pairs, the 
Feynman diagram of which is also shown in Fig.~\ref{fig1} (right). Two-photon interactions and 
coherent photoproduction are both associated with a low $p_T$ ($< \; \approx$100~MeV/c) of the 
final state. To take into account also the finite detector resolution, a cut of $p_T <$~300~MeV/c 
is used to define coherent interactions in the present analysis. 

The ALICE detector and the 
muon spectrometer are discussed in section 2. Section 3 describes the data analysis and the 
cross section measurement. In section 4, finally, the cross section is compared with model 
expectations.

\section{The ALICE Experiment}

The ALICE detector consists of a central barrel placed inside a large solenoid magnet covering 
the pseudorapidity range $| \eta | <$~0.9, a muon spectrometer covering the range 
$-4.0 < \eta < -2.5$, and a set of smaller detectors at forward rapidities. The current analysis 
is based on data from the muon spectrometer. In addition, the VZERO counters and Zero-Degree Calorimeters 
(ZDC) are used for triggering and rejecting the contribution from hadronic interactions. 

The ALICE muon spectrometer consists of five tracking stations containing two planes of cathode pad 
multi-wire proportional chambers. A ten interaction length thick absorber is placed 
between the primary vertex and the first tracking station. The third (middle) tracking station is 
situated inside a dipole magnet with a $\int B dl =$~3~Tm integrated field. The muon spectrometer 
also includes a triggering system consisting of four planes of 
resistive plate chambers, which are placed behind a seven interaction length thick iron wall. 
The iron wall absorbs secondary punch through hadrons from the front absorber and low momentum 
muons from $\pi$ and K weak decays. 

The VZERO counters are arrays of scintillator tiles situated on either side of the primary vertex 
at pseudorapidities $2.8 < \eta < 5.1$ (VZERO-A, on the opposite side of the muon spectrometer) and 
$-3.7 < \eta < -1.7$ (VZERO-C, on the same side as the muon spectrometer). Information from the VZERO 
is available at the lowest trigger level in ALICE. 

The ZDCs are hadronic calorimeters located at 116 m on either side of the interaction point. They detect 
neutrons in the very forward region ($|\eta| >$~8.8). 

The present analysis is based on a sample of $3.16 \times 10^6$ events collected with a special 
trigger for ultra-peripheral collisions in the forward region (FUPC) during the 2011 Pb-Pb run. The 
purpose of the FUPC trigger was to 
select two muons in an otherwise empty detector. It is based on three requirements: 
a single muon trigger 
above a 1 GeV/c $p_T$--threshold; at least one hit in VZERO-C; no hits in VZERO-A. The 
integrated luminosity for the data collected with this trigger in 2011 corresponds to 
about $55 \mu b^{-1}$.

\section{Data analysis}

The offline event selection was done in such a way as to maximize the yield of exclusively 
produced muon pairs from $J/\psi$ decay and two-photon interactions, while minimizing the 
background from hadronic collisions and beam gas interactions. Only events with two 
oppositely charged tracks in the muon spectrometer were considered. The analysis was restricted to 
$J/\psi$ rapidities between $-3.6 < y < -2.6$ and muon pseudorapidities between 
$-3.7 < \eta_{1,2} < -2.5$ to match the acceptance of the VZERO-C and avoid 
the edges of the spectrometer. A $p_T$ dependent cut on the distance of closest approach 
of the tracks from the primary vertex position in the transverse plane was applied, and at 
least one of the muon track candidates had to match a trigger track above the 1~GeV/c 
threshold. More details about the analysis cuts are available in~\cite{Abelev:2012ba}. 

Exclusive photoproduction of vector mesons normally leaves the nuclei intact. But the strong fields 
associated with heavy-ions at high energies make exchange of multiple photons possible. 
These additional photons have low energy but may lead to nuclear break up, followed by emission 
of one or a few neutrons in the forward region. The energy deposit in the ZDCs was therefore not required 
to be zero but only to be less than 6 TeV. 
This cut reduces the hadronic background at higher $J/\psi$ transverse momentum but does 
not remove any events with $p_T < 0.3$~GeV/c. 

\begin{figure}
\includegraphics[height=4.5cm]{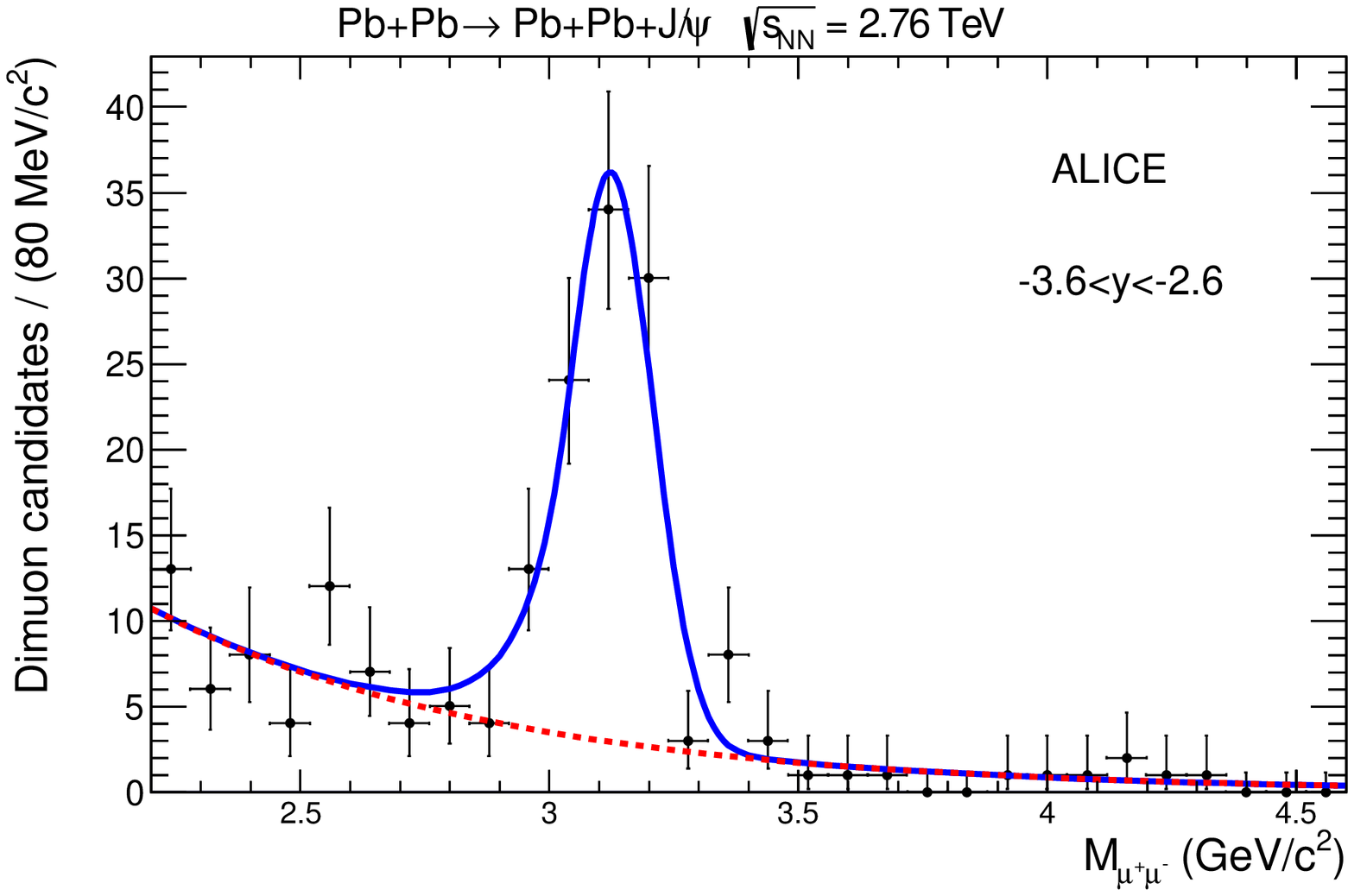}
\includegraphics[height=4.5cm]{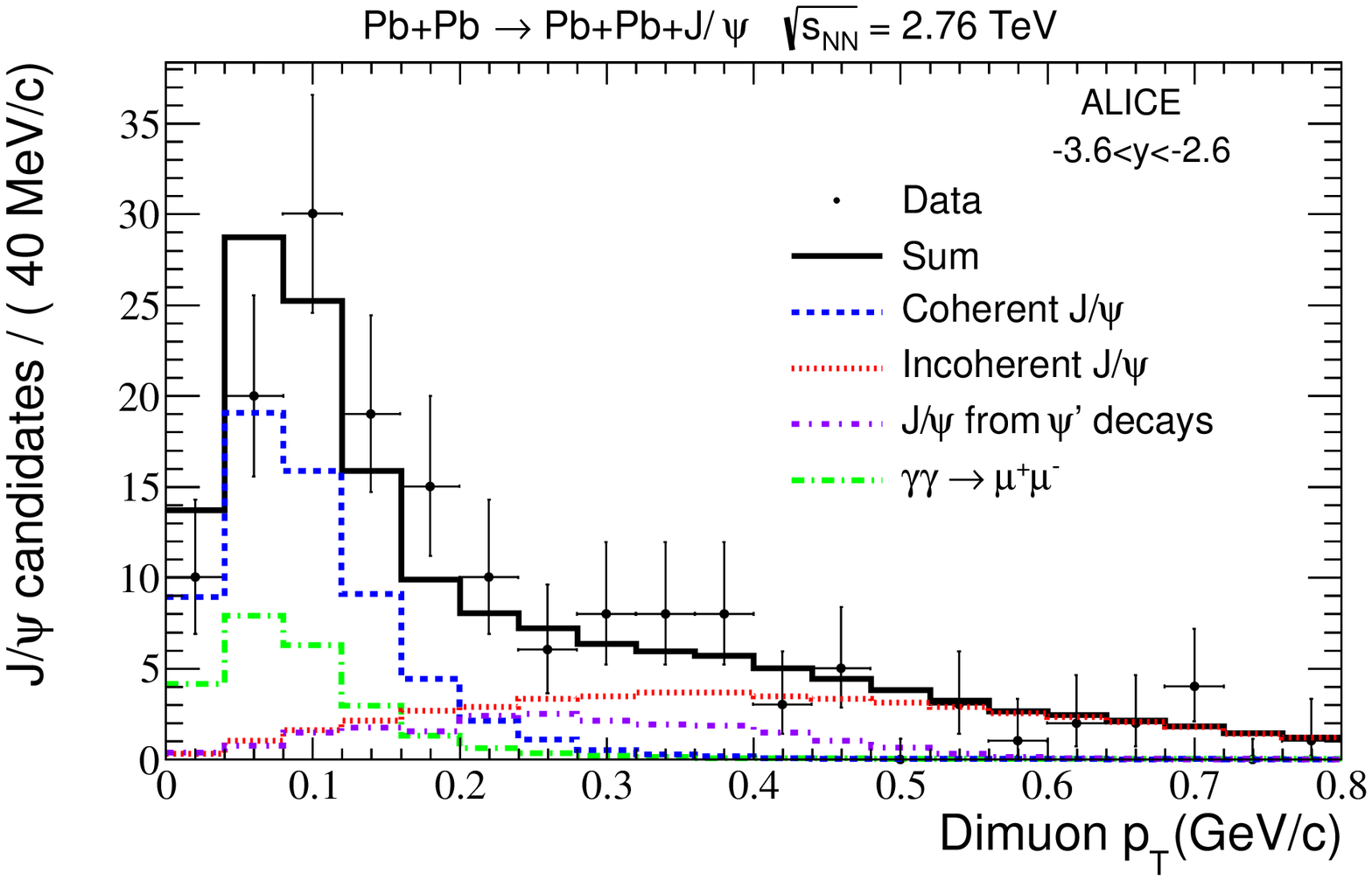}
\caption{Invariant mass distribution (left) for dimuons surviving the analysis cuts. The curve shows 
the sum of a Crystal Ball function and an exponential fitted to the data, as described in the text. 
Transverese momentum distribution (right) for $J/\psi$ candidates with $2.8 < m_{inv} < 3.4$~GeV/c$^2$. 
The histograms are explained in the text.}
\label{fig2}
\end{figure}

The invariant mass distribution for the dimuons surviving the analysis cuts 
are shown in Figure~\ref{fig2} (left). 
The final sample contained 117 $J/\psi$ candidates in the invariant mass 
range $2.8 < m_{inv} < 3.4$~GeV/c$^2$. The number of $J/\psi$s was extracted by fitting 
the invariant mass distribution in Fig.~\ref{fig2} to the sum of a Crystal Ball function, representing the 
signal, and an exponential, representing the background from two-photon interactions. 
This gave an extracted number of $J/\psi$s $N_{yield} = 96 \pm 12 (\rm{stat}) \pm 6 (\rm{syst})$. 
The systematic error on the yield was obtained by varying the Crystal Ball tail parameters. 

The transverse momentum distribution of $J/\psi$ candidates is shown in Fig~\ref{fig2} (right).  
The histograms show the expected contribution from coherent $J/\psi$ production, 
incoherent $J/\psi$ production, $J/\psi$ from the decay $\psi' \rightarrow J/\psi + X$ , and 
two-photon production of dimuon pairs. As can be seen, the signal region $p_T <$~0.3~GeV/c 
contains background contributions from incoherent $J/\psi$ production as well as feed down from
$\psi'$ decay.

These contributions can be expressed as fractions of the number of 
coherent $J/\psi$s ($N_{J/\psi}^{coh}$):
\begin{equation}
\begin{array}{cc}
f_I = \frac{N_{J/\psi}^{incoh}}{N_{J/\psi}^{coh}} \; , & 
f_D = \frac{N_{\rm{FD}}}{N_{J/\psi}^{coh}} \\
\end{array}
\end{equation} 


The method used to estimate these fractions and the associated uncertainties is 
explained in detail in~\cite{Abelev:2012ba}. The fraction of $J/\psi$s from feed-down was 
obtained from the calculated  $\psi'$ photoproduction cross section from the models 
in~\cite{starlight} and~\cite{Rebyakova:2011vf}. The acceptance and efficiency for reconstructing a $J/\psi$ was 
obtained by simulating the decay $\psi' \rightarrow J/\psi + X$ 
with Pythia and passing the events through the ALICE Geant detector simulation package. 
The result was $f_D = 0.11 \pm 0.06$. 

The fraction of incoherent events was similarly obtained from model 
calculations~\cite{starlight,Rebyakova:2011vf} followed by simulations of the detector response. It 
was also, as a cross check, obtained by fitting the $p_T$ distribution in Figure~\ref{fig2} to 
Monte Carlo templates representing the four contributions mentioned above. The relative normalization 
for coherent and incoherent production was left free 
in the fit, while the contribution from feed-down from photoproduced $\psi'$ was constrained from the 
estimate $f_D = 0.11 \pm 0.06$. The two-photon contribution was obtained from the fit of the dimuon 
continuum outside the $J/\psi$ peak. This gave the result $f_I = 0.12 ^{+0.14}_{-0.04}$. 

The contribution from hadronic $J/\psi$ production was estimated from the measured yield above 
$p_T >$~1~GeV/c, where the contribution from photoproduction is negligible. It was also estimated 
by scaling the measured $J/\psi$ yield in p-p collisions to the 20\% most peripheral, hadronic 
Pb-Pb collisions. Both estimates showed that the contribution from hadronic production is negligible 
in the $p_T <$~0.3~GeV/c region considered here. 

The number of coherent $J/\psi$s is thus related to the extracted yield by 
\begin{equation}
N_{J/\psi}^{coh} = \frac{N_{yield}}{1 + f_I + f_D} \; .
\end{equation} 
The resulting 
number of coherent $J/\psi$s is $N_{J/\psi}^{coh} = 78 \pm 10 (\rm{stat}) ^{+7}_{-11}(\rm{syst})$. 

During the 2011 Pb-Pb run, the VZERO detector had a threshold corresponding to an energy deposit 
above that from a minimum ionizing particle. This made it difficult to accurately simulate the 
VZERO-C trigger efficiency for low multiplicity events. 
To avoid these uncertainties, the 
$J/\psi$ cross section was obtained by using the number of reconstructed 
$\gamma \gamma \rightarrow \mu^+ \mu^-$ events. Two photon production of dimuon pairs is a 
standard QED process which has been proposed earlier as a luminosity monitor~\cite{Baur:2001jj}. The 
$J/\psi$ cross section can then be written in a way that is independent of the trigger efficiency 
and the integrated luminosity 
\begin{equation}
\frac{\mathrm{d}\sigma_{J/\psi}^{\mathrm{coh}}}{\mathrm{d} y}   =
\frac{1}{BR(J/\psi \rightarrow \mu^+ \mu^-)}
\cdot
\frac{N^{\mathrm{coh}}_{J/\psi}}{N_{\gamma\gamma}}
\cdot
\frac{(\mathrm {Acc}\times\varepsilon)_{\gamma\gamma} }
{(\mathrm{Acc}\times\varepsilon)_{J/\psi}}
\cdot
\frac{\sigma_{\gamma\gamma}}{\Delta y} \; .
\label{comp2}
\end{equation}
Here, $(\mathrm{Acc}\times\varepsilon)_{\gamma\gamma}$ and $(\mathrm{Acc}\times\varepsilon)_{J/\psi}$ are 
the acceptance and efficiency of the muon spectrometer for two-photon and coherent $J/\psi$ events, 
respectively. $BR(J/\psi \rightarrow \mu^+ \mu^-) =$~5.93\% is the branching ratio for the dimuon 
decay of the $J/\psi$ and $\Delta y =$~1.0 is the rapidity interval. The cross section $\sigma_{\gamma \gamma}$ 
is calculated using STARLIGHT for a dimuon within $-3.6 < y < -2.6$ and each muon satisfying 
$-3.7 < \eta_{1,2} < -2.5$. The cross section is calculated for two intervals in invariant mass on each 
side of the $J/\psi$ peak ($2.2 < m_{inv} < 2.6$~GeV/c$^2$ and $3.5 < m_{inv} < 6.0$~GeV/c$^2$). 
The number of events found in each interval is 43 and 15, for the low and high mass interval, 
respectively. This together with the cross sections, $\sigma_{\gamma \gamma} =$13.7~$\mu$b (low $m_{inv}$) 
and 3.7~$\mu$b (high $m_{inv}$), gave a total cross section 
\begin{displaymath}
\frac{\mathrm{d}\sigma_{J/\psi}^{\mathrm{coh}}}{\mathrm{d} y} = 1.00 \pm  0.18 (\mathrm{stat}) 
^{+0.24}_{-0.26} (\mathrm{syst}) \, \mathrm{mb} .
\end{displaymath}
The systematic error is dominated by the uncertainty in the two-photon cross secion. Although this is a 
standard QED process, the fact that the coupling to the nuclei is not small ($Z \sqrt{\alpha}$ rather than
$\sqrt{\alpha}$) and that the photon emitting nuclei are extended objects with an internal structure introduces 
a significant uncertainty in the cross section. Based on calculations~\cite{Baltz:2009fs} and 
constraints from data from RHIC~\cite{Afanasiev:2009hy,Adams:2004rz}, this 
uncertainty is estimated to be 20\%. Other important contributions to the systematic error are the 
coherent signal extraction and the reconstruction efficiency~\cite{Abelev:2012ba}.

\section{Comparison with models}

The measured cross section can be compared with the models mentioned 
earlier~\cite{starlight,Lappi:2013am,Rebyakova:2011vf,Adeluyi:2012ph,Cisek:2012yt,Goncalves:2011vf}. 
The models by Lappi and M{\"a}ntysaari~\cite{Lappi:2013am}, Cisek, Sch{\"a}fer and Szczurek~\cite{Cisek:2012yt}, 
and Goncalves and Machado~\cite{Goncalves:2011vf} are based on the Color Dipole Model (CDM). The 
model by Klein and Nystrand\cite{starlight}, which is incorporated in the STARLIGHT Monte Carlo, uses data from 
exclusive vector meson production at HERA as input to a Glauber calculation of the cross section 
for nuclear targets. The models by Rebyakova, Strikman and Zhalov~\cite{Rebyakova:2011vf}, and 
Adeluyi and Bertulani~\cite{Adeluyi:2012ph}, calculate the cross section directly from the nuclear 
gluon distribution with the forward scattering amplitude being proportional to the gluon 
distribution squared. Rebyakova, Strikman and Zhalov calculate the modifications to the nuclear 
gluon distrbution in the leading twist approximation, while Adeluyi and Bertulani use some of the 
standard parameterizations (HKN07, EPS09, and EPS08). Adeluyi and Bertulani also calculate the 
cross section by scaling the $\gamma + p \rightarrow J/\psi +p$ cross section with the number of 
nucleons assuming no nuclear effects (MSTW08). 

\begin{figure}
\includegraphics[height=5.0cm]{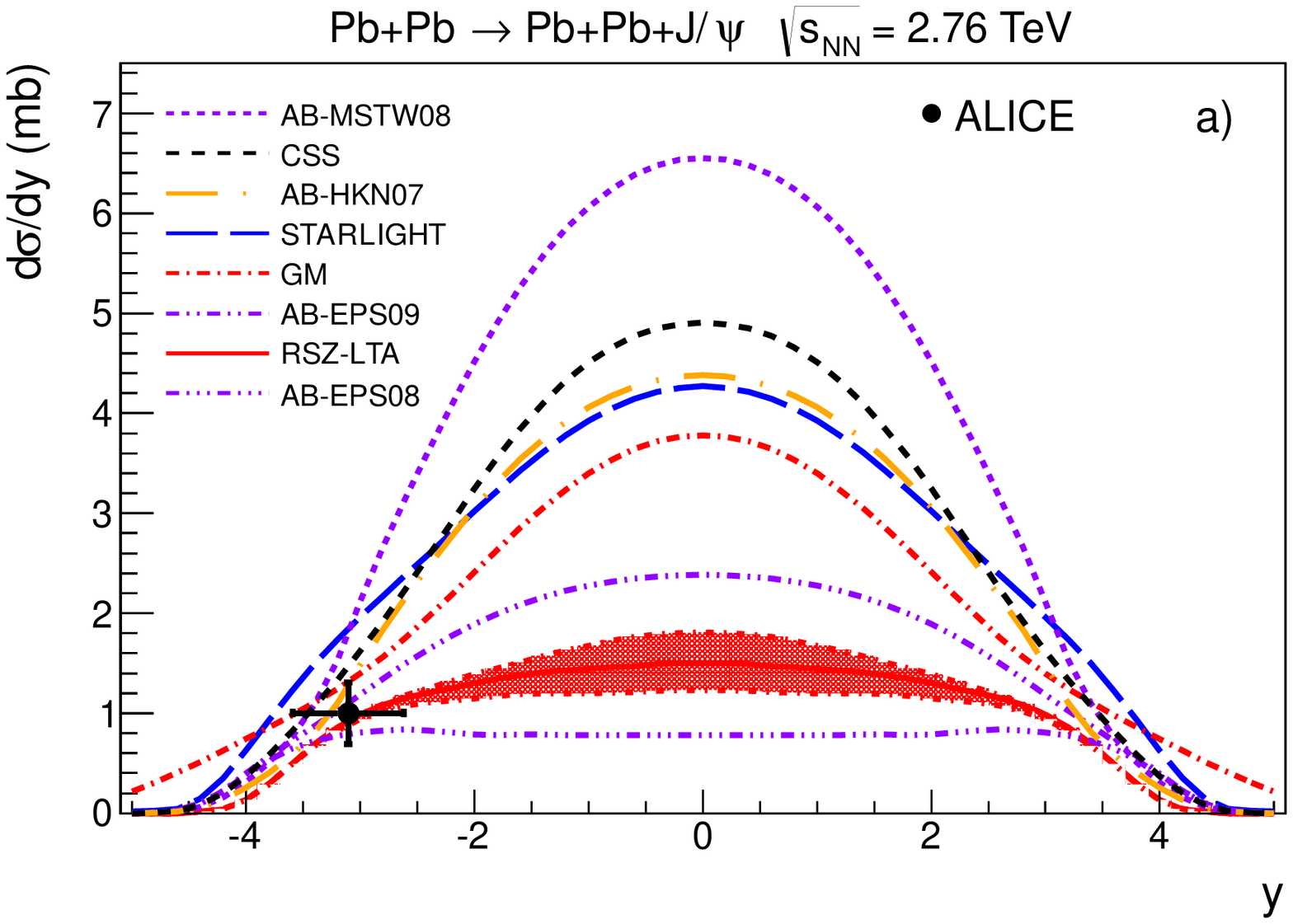}
\includegraphics[height=5.0cm]{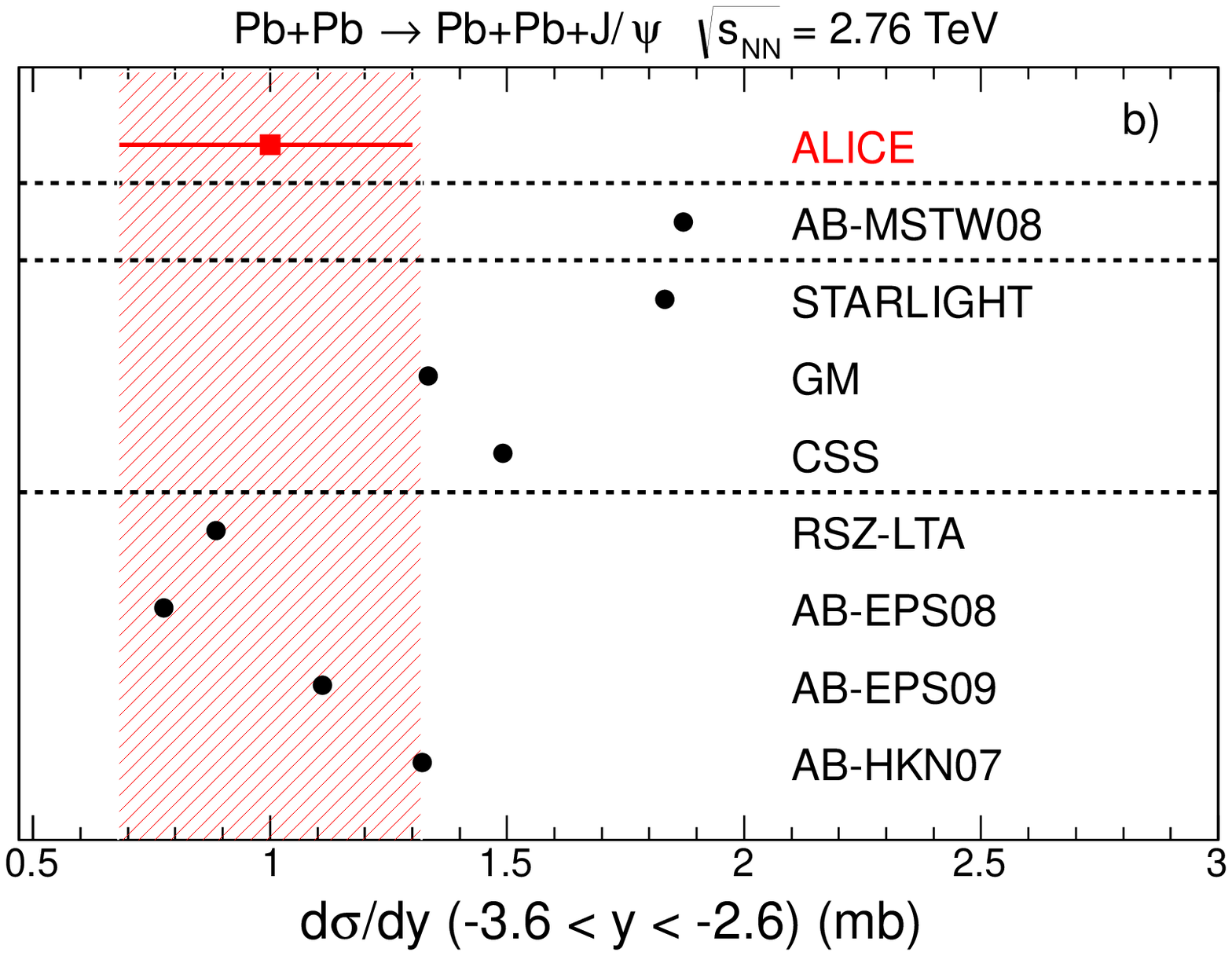}
\caption{The cross section from ALICE compared with $d\sigma/dy$ from models (left), and a comparison 
of the cross section integrated over $-3.6 < y < -2.6$ (right). The experimental error is the squared 
sum of the statistical and systematic errors.}
\label{fig:models}
\end{figure}

The results are shown in Fig.~\ref{fig:models}. One can see that models which are based on the color dipole 
model generally give a higher cross section than those which calculate the cross section directly from 
the gluon distribution. This is true at the forward rapidities studied here, but is emphasized even more 
at mid-rapidity. The MSTW08 scaling of the cross section without nuclear effects and STARLIGHT deviate by 
about 3 standard deviations from the measured value and are disfavored. Best agreement is found for models 
which include nuclear gluon shadowing consistent with the EPS09 or EPS08 parameterizations. The calculation by 
Lappi and M{\"a}ntysaari is in the same range as the other models using the color dipole model, but since 
their result was not available when the ALICE paper was released it is not included in the figure.

\section{Conclusions and outlook}

The cross section for exclusive $J/\psi$ production in Pb-Pb collisions at the LHC has been measured 
by the ALICE collaboration. The results show that the cross section cannot be undestood from a simple 
scaling of the nucleon cross section neglecting nuclear effects. Best agreement is seen with models which 
include nuclear gluon shadowing. The cross section for exclusive $J/\psi$ production at mid-rapidity is 
being studied in ALICE and final results will be published shortly.


\begin{thebibliography}{99}

\bibitem{Review2008}
A.~J.~Baltz {\it et al.}
Phys. Rept. {\bf 458} (2008) 1.
%

\bibitem{Review2005}
C.~A.~Bertulani, S.~R.~Klein and J.~Nystrand,
Ann. Rev. Nucl. Part. Sci. {\bf 55} (2005) 271.
%

\bibitem{Abelev:2012ba}
  B.~Abelev {\it et al.}  [ALICE Collaboration],
  Phys.\ Lett.\ B {\bf 718} (2013) 1273.

\bibitem{starlight}
S.~R.~Klein and J.~Nystrand,
Phys. Rev. C {\bf 60}, 014903 (1999).
%

\bibitem{Lappi:2013am}
  T.~Lappi and H.~M{\"a}ntysaari,
  arXiv:1301.4095 [hep-ph].

\bibitem{Rebyakova:2011vf}
  V.~Rebyakova, M.~Strikman and M.~Zhalov,
  Phys.\ Lett.\  B {\bf 710} (2012) 647.

\bibitem{Adeluyi:2012ph}
  A.~Adeluyi and C.~A.~Bertulani,
  Phys.\ Rev.\ C {\bf 85} (2012) 044904.

\bibitem{Cisek:2012yt}
  A.~Cisek, W.~Sch{\"a}fer and A.~Szczurek,
  Phys.\ Rev.\ C {\bf 86} (2012) 014905.

\bibitem{Goncalves:2011vf}
  V.~P.~Goncalves and M.~V.~T.~Machado,
  Phys.\ Rev.\ C {\bf 84} (2011) 011902.

\bibitem{Adler:2002sc}
  C.~Adler {\it et al.}  [STAR Collaboration],
  Phys.\ Rev.\ Lett.\  {\bf 89} (2002) 272302.

\bibitem{Afanasiev:2009hy}
  S.~Afanasiev {\it et al.}  [PHENIX Collaboration],
  Phys.\ Lett.\  B {\bf 679} (2009) 321.

\bibitem{Baur:2001jj}
  G.~Baur, K.~Hencken, D.~Trautmann, S.~Sadovsky and Y.~Kharlov,
  Phys.\ Rept.\  {\bf 364} (2002) 359. 

\bibitem{Baltz:2009fs}
  A.~J.~Baltz,
  Phys.\ Rev.\ C {\bf 80} (2009) 034901.

\bibitem{Adams:2004rz}
  J.~Adams {\it et al.}  [STAR Collaboration],
  Phys.\ Rev.\ C {\bf 70} (2004) 031902.

\end{thebibliography}
\end{document}